\documentclass[aps,pra,reprint,showpacs,amssymb,amsmath]{revtex4-1}

\usepackage{graphicx}
\usepackage{times}
\usepackage{color}
\usepackage{siunitx}

\newcommand{\Erec}{E_\mathrm{R}}

\newcommand{\qmax}{q_\mathrm{max}}
\newcommand{\kr}{k_\mathrm{R}}
\newcommand{\llat}{\lambda_\mathrm{Lat}}
\newcommand{\Vlat}{V_\mathrm{Lat}}

\newcommand{\waxial}{\omega_\mathrm{a}}
\newcommand{\wradial}{\omega_\mathrm{r}}

\newcommand{\numod}{\nu_\mathrm{mod}}
\newcommand{\nudeex}{\nu_\mathrm{de-ex}}
\newcommand{\tref}{t_\mathrm{ref}}
\newcommand{\Tref}{\tau_\mathrm{ref}}

\newcommand{\Rb}{$^{87}$Rb }
\newcommand{\Rbnospace}{$^{87}$Rb}
\newcommand{\rme}{\mathrm{e}}
\newcommand{\braopket}[3]{\langle#1|#2|#3\rangle}

\begin{document}
\title{Production and manipulation of wave packets from ultracold atoms in an optical lattice}
\author{Poul L. Pedersen}\email[]{poul@phys.au.dk}
\author{Miroslav Gajdacz}
\author{Nils Winter}
\author{Andrew J. Hilliard}
\author{Jacob F. Sherson}
\author{Jan Arlt}
\affiliation{Danish National Research Foundation Center for Quantum Optics, Institut for Fysik og Astronomi, Aarhus Universitet, Ny Munkegade 120, 8000 Aarhus C, Denmark.}

\date{\today}

\begin{abstract}
Within the combined potential of an optical lattice and a harmonic magnetic trap, it is possible to form matter wave packets by intensity modulation of the lattice. An analysis of the production and motion of these wave packets provides a detailed understanding of the dynamical evolution of the system. The modulation technique also allows for a controllable transfer (de-excitation) of atoms from such wave packets to a state bound by the lattice. Thus, it acts as a beam splitter for matter waves that can selectively address different bands, enabling the preparation of atoms in  localized states. The combination of wave packet creation and de-excitation closely resembles the well-known method of pump-probe spectroscopy. Here, we use the de-excitation  for  spectroscopy of the anharmonicity of the combined potential. Finally, we demonstrate that lattice modulation can be used to excite matter wave packets to even higher momenta, producing fast wave packets with potential applications in precision measurements. 
\end{abstract}
\pagestyle{plain}
\maketitle

\section{Introduction}

Within the past two decades, rapid progress in the field of cold gases has been driven by the ability to manipulate these ensembles with unprecedented precision. This has led to a multitude of results, ranging from fundamentally new experiments on quantum systems~\cite{Bloch2008} to improved precision measurements~\cite{vanZoest18062010,PhysRevLett.97.240801}. In particular, the confinement and manipulation of ultracold atoms in optical lattices  has led to the new research area of quantum simulation of solid state systems~\cite{Leweenstein_Optlattreview}. 

One intriguing current direction is the investigation of Hubbard models with an extended interaction range, which can either be realized using dipolar gases~\cite{0034-4885-72-12-126401} or atoms in higher bands interacting through zero-range collisions~\cite{PhysRevLett.95.033003}. For the latter, the central challenge consists in the high fidelity preparation of atoms in selected bands. Experimental steps in this direction have been taken using Raman transfer \cite{Muller2007}, dynamical control of the potential~\cite{DeChiara:2008,Wirth2010}, and  Bragg spectroscopy~\cite{1367-2630-11-10-103030}. 

The combined potential of a magnetic trap and an optical lattice allows one to address individual bands   using both bosons~\cite{Ott-2004} and fermions~\cite{Heinze13}, and it was previously shown in our experiments that wave packets can be used to populate individual states in this system with high efficiency~\cite{Sherson2012}. These wave packets are created by lattice intensity modulation and evolve in a quasi-continuum of states bound by an external magnetic trap. Their de-excitation into lower bands to populate  localized states can then serve as a resource for quantum  simulation.

Since the term `wave packets' is used in a number of different contexts in the field of ultracold atoms, we limit the following brief discussion of other experiments to cases where optical lattices are used to manipulate Bose-Einstein condensates (BECs)~\cite{Morsch2006}. Wave packets can be created by exposing  a BEC to brief pulses of optical lattice light to split the cloud. This method can generate wave packets propagating in free space~\cite{Kozuma-1999}, in a harmonic trap~\cite{Denschlag-2000}, or in wave guides~\cite{Fabre2011,*[{See }][{  for related work on ultracold thermal clouds.}], Cheiney13}. Additionally, atoms performing Bloch oscillations in the lowest band form wave packets in higher bands through Landau-Zener transitions ~\cite{Morsch2001,Alberti2009,Haller2010,Park2012}. 

The lattice modulation technique employed here has previously been used to create quasi-free wave packets in the combined potential of an optical lattice and a magnetic trap~\cite{Sherson2012}. It was also shown that  these wave packets can be de-excited into bound lattice bands. However, a detailed analysis of the wave packet evolution and the maximal efficiency of the excitation process was limited by the available data quality.

Within this work, we characterize  the creation of wave packets and their motion in detail, revealing previously unobserved features. Moreover,  the de-excitation technique is extended to all available bands and it is shown to operate with an efficiency approaching unity. We also demonstrate that it can serve as a tool for  spectroscopy of the confining potential~\cite{PhysRevA.75.043602}.
  In a further extension of the wave packet technique, we create and characterize  wave packets in higher quasi-continuum states at large velocities.
  
The creation and de-excitation process strongly resembles the well-established technique of pump-probe spectroscopy, 
 where an initial pump pulse creates an excitation in the system, which is interrogated by a probe pulse at a later time,  transferring the excited state population to a chosen final state. 
 We  exploit this analogy by showing that the technique provides a high degree of control over transfer to the vibrational bands of the lattice. This has prospective applications in quantum metrology, complementing previous work on large momentum beam splitters~\cite{Denschlag2002,PhysRevLett.100.180405,Clade2009,Park2012}.

This paper is organized as follows: In section~\ref{sec:exp-wp}, we briefly describe the experimental setup and the creation of wave packets in our system. A simple band model is presented, and we show that this model describes the observed wave packet motion accurately. In section~\ref{sec:wp-motion}, the wave packet motion is characterized, and the parameters available for controlling the motion are explored. In section~\ref{sec:deexcitation}, the effect of varying different modulation parameters is investigated along with their effect on de-exciting a wave packet into a localized state. Section~\ref{sec:up-excitation} presents more experiments to probe the band structure in momentum space.

\section{Experimental realization of wave packets}\label{sec:exp-wp}
\begin{figure}[t]
  \includegraphics[width=8.6cm]{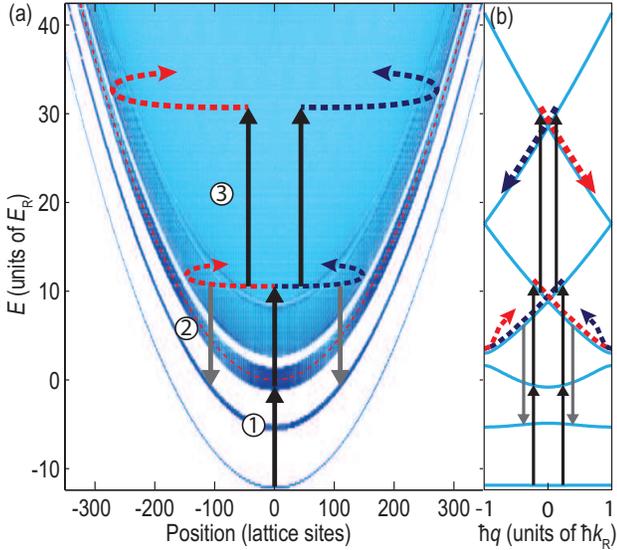}
  \caption{(Color online) (a) Calculated spectrum of the 1D single particle Schr\"{o}dinger equation for the combined magnetic trap and optical lattice potential. Each horizontal line represents the summed population density of the eigenstates within an energy bin of width $h\cdot664$~Hz. 
  The shading becomes darker with increasing density of states and white indicates forbidden regions. The dashed red line indicates the local maximum of the combined potential. Three experimental processes are shown: (1) A wave packet can be generated via a two-step excitation process. (2) A wave packet can be de-excited into  lower lying localized states. (3) A wave packet can be excited to higher states. (b) The same processes are shown in the conventional band picture represented in the first Brillouin zone versus quasi-momentum $\hbar q$.}
  \label{fig:spectrum}
\end{figure}

Wave packets of ultracold atoms can be created in the combined potential of a one dimensional optical lattice and a Ioffe-Pritchard magnetic trap by intensity modulation of the lattice. The creation of these wave packets was previously described in Ref.~\cite{Sherson2012} and is only briefly repeated here.
 Figure~\ref{fig:spectrum} shows a single particle calculation of the system's band structure.
 For our choice of parameters, three energy bands are bound by the lattice, and a region of densely spaced states exists above the second excited band. The spacing between adjacent energy levels in this region is $\sim 30$~Hz and may be regarded as comprising a quasi-continuum in comparison with the $\sim10$~kHz spacing between energy bands.
By modulating the lattice intensity at an appropriate frequency,
various transitions between the bands and the quasi-continuum can be driven to create, excite or de-excite wave packets  (see Fig.~\ref{fig:spectrum}).

To initiate these experiments, a BEC of $10^5$ \Rb atoms in the $5^2 S_{1/2} |F=2, m_F=2\rangle$ state is adiabatically transferred into a potential consisting of a magnetic trap with axial and radial trapping frequencies of $\waxial = 2\pi\times 12.2$~Hz and \mbox{$\wradial = 2\pi\times 40.6$~Hz}, and an optical lattice at a wavelength of 
 \mbox{$\llat = \SI{914}{\nm}$}. 
 We use a lattice depth of \mbox{$s \equiv\Vlat/\Erec = 16$}, where $\Vlat$ is the depth of the potential and $\Erec = (\hbar\kr)^2/(2m)=\hbar(2\pi\times2.75$~kHz) is the recoil energy, with \mbox{$\kr=2\pi/\llat$} and $m$ the mass of \Rbnospace. 

To produce wave packets, we modulate the depth of the lattice by varying the lattice beam intensity \cite{Denschlag2002} according to
\begin{equation}
V(y,t)=\frac{1}{2}m\wradial^2y^2-s\Erec\cos^2(\kr y)(1+\epsilon\cos(2\pi\numod t)),
\end{equation}
where the strength of the modulation is characterized by the relative amplitude $\epsilon$. The production of wave packets is constrained by this method of excitation since the matrix element linking initial and final states is proportional to: $\braopket{n_i}{\cos^2(\kr y)}{n_f}=\braopket{n_i}{1-(\kr y)^2+(\kr y)^4/12+\dots}{n_f}$. To lowest order, this leads to the selection rule $\Delta n = 2$~\cite{PhysRevA.56.R1095}.

The excitation of wave packets into the quasi-continuum is a two step process: The first excitation step takes the atoms from the lowest band to the second excited band, and the second step excites the atoms into the quasi-continuum at an energy corresponding to the fourth excited band as detailed in Ref.~\cite{Sherson2012} and illustrated in Fig.~\ref{fig:spectrum} (process (1)). The atoms populate a superposition of eigenstates in the quasi-continuum, corresponding to two wave packets of opposite momentum  since the band structure is symmetric in momentum space. Hence, one wave packet initially moves upwards, and the other moves downwards~\footnote{We refer to these as the upper and lower wave packets} in the combined potential. 
During the subsequent movement, the wave packets exchange their initial kinetic energy for potential energy by moving outward in the harmonic potential while `rolling down' the appropriate bands~\footnote{While the wave packet moves between the third and fourth excited bands, it encounters a small perturbation which, for an infinite system, would correspond to a band gap of $0.20~\Erec$. However, in the real system, it appears as a reduction in the density of states. This reduction is small in the sense that the density of states between the third and fourth excited bands decreases by an order of magnitude, whereas between the second and third excited bands the decrease is more than 15 orders of magnitude. We have not observed any effect of this small gap on wave packet dynamics.}.
In order to gain spatial information on the wave packets, we measure the number of atoms by in-trap absorption imaging \footnote{ The images are typically taken in a decompressed magnetic trap with a bias field of 11~G, such that each of the magnetic sublevels is Zeeman shifted, thereby broadening the resonance. We take this into account by calibrating the in-trap imaging against standard absorption imaging for identically prepared cold atomic clouds. The atom number obtained from absorption imaging was calibrated following the method described in~\cite{Mirek}. The combined potential is sufficiently weak that the imaging is independent of the wave packet position.}.

\begin{figure}[]
  \includegraphics[width=8.6cm]{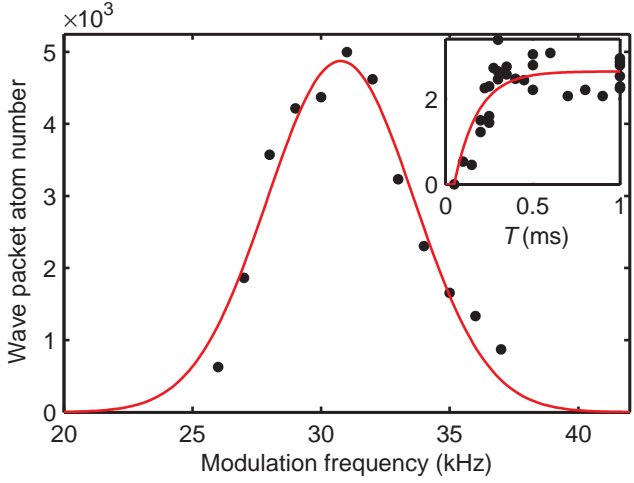}
	\caption{(Color online) Number of atoms in the lower wave packet as a function of the modulation frequency for a  pulse of  $T=500~\mu$s duration. The data was fitted with a Gaussian as an empirical model yielding an optimal transfer frequency of $\numod=$30.8~kHz, and a $1/\rme^2$ width of 5.7~kHz. Inset: Number of atoms in the lower wave packet as a function of modulation duration $T$ for $\numod=$33~kHz (same vertical axis units as main figure); the data was fitted with an exponential saturation function with time constant $\tau=130~\mu$s. For both data sets, the relative modulation amplitude was $\epsilon=0.165$.}
	\label{fig:pump-vs-freq}
\end{figure}
In the following, we investigate the wave packet production efficiency  as a function of modulation frequency and duration. This data underpins the choice of parameters in Ref.~\cite{Sherson2012} and the rest of this work. The number of atoms transferred to the wave packets was measured for different modulation frequencies;  Fig.~\ref{fig:pump-vs-freq} shows the data  for the lower wave packet. The data was obtained for $T=500~\mu$s  modulation duration and a relative modulation amplitude $\epsilon=0.165$, with a rectangular amplitude envelope. The frequency that produced the largest transfer was 30.8~kHz; this agrees well with the $\sim 11\Erec/\hbar$ band spacing evident in Fig.~\ref{fig:spectrum}. The resonance was fitted with a Gaussian as an empirical model, yielding a $1/\rme^2$ width of 5.7~kHz; this measured width is consistent with the 7~kHz width of the second excited band. 

The inset to Fig.~\ref{fig:pump-vs-freq} shows the number of atoms in the lower wave packet as a function of time for a modulation frequency of 33~kHz and relative modulation amplitude $\epsilon=0.165$. The data was fitted with an exponential saturation function ($\propto[1-\exp(-t/\tau)]$) to characterize the time required to saturate the transfer, yielding the time constant $\tau=130~\mu$s~\footnote{A time offset was included in the fitting function to allow for a redistribution of the energy of the atoms in the second excited band. This is necessary to meet the resonance condition to the quasi-continuum as shown in Fig.~\ref{fig:spectrum}~(b).}. Due to the large number of excited states in the second excited band and the quasi-continuum, we do not observe any coherent oscillations of the population. The transfer is nearly saturated at $T=500~\mu$s, and we therefore use amplitude modulation pulses with this duration in the following.

This characterization of the initial wave packet creation provides the starting point for an analysis of the wave packet motion and for further de-excitation and excitation studies.

\section{Wave packet motion}\label{sec:wp-motion}

\begin{figure}[t]
	 \includegraphics[width=8.59cm]{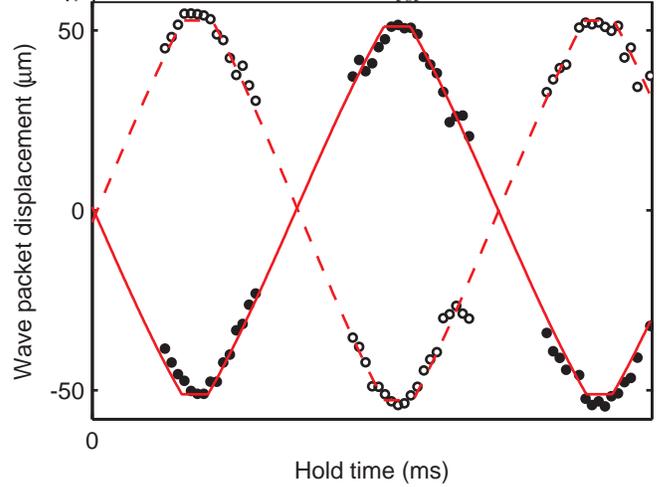}
	\caption{(Color online) Wave packet motion in the quasi-continuum following modulation at 30~kHz, fitted with Eqn.~\eqref{eq:cutsine}. Lower wave packet data and fit: filled circles, solid line; upper wave packet data and fit: open circles, dashed line. The wave packets oscillate harmonically but with  turning regions caused by reflections at the energy gap between the second and third excited bands.}
	\label{fig:wp-move}
\end{figure}
The wave packet evolution in the combined potential of the magnetic trap and the optical lattice was initially analyzed in Ref.~\cite{Sherson2012}. %
Each wave packet undergoes oscillatory motion due to the harmonic confinement of the magnetic trap until it reaches 
 the gap between the second and third excited bands. At this gap ($q = 3\kr$), it encounters a potential barrier and, depending on the size of the band gap, one of two things can happen: the atoms can undergo Landau-Zener tunneling, or they follow the avoided crossing into the next Brillouin zone, corresponding  to an inversion of the momentum, i.e. Bragg reflection.  In our system, the vast majority of the atoms undergo Bragg reflection since the Landau-Zener tunneling probability is 0.1\% \cite{zener32}. 
 Within previous work~\cite{Sherson2012}, the detailed features of the oscillation were not resolved.
 
Figure~\ref{fig:wp-move} shows the  motion of the upper and lower wave packets in the quasi-continuum following $500~\mu$s modulation at 30~kHz. 
The wave packet is reflected at time $\tref$ and after traveling through the main cloud, it is reflected again when it reaches the opposite side of the trap at time $3\tref$. These reflections are repeated at odd multiples of $\tref$. To account for the finite size of each wave packet, the reflection at the band gap is modeled to take time $\Tref$. 
We thus model the displacement of each wave packet by
\begin{widetext}
\begin{equation}
y = \begin{cases}
A\sin(\wradial t+\phi_0) & 0 \leq t < \tref- \frac{\Tref}{2}\\
A\sin(\wradial(\tref-\Tref/2)+\phi_0) &  \tref- \frac{\Tref}{2} \leq t < \tref+ \frac{\Tref}{2}\\
A\sin\left(\wradial t + \phi_0+\phi_\text{ref}\right) & \tref+\frac{\Tref}{2}\leq t < 3\tref-\frac{\Tref}{2}\\
\dots
\end{cases}
\label{eq:cutsine}
\end{equation}
\end{widetext}
where $\phi_\text{ref} = \pi-2\wradial\tref$ is the phase shift due to Bragg reflection. 
 
The displacement of each wave packet in Fig.~\ref{fig:wp-move} was fitted with Eqn.~\eqref{eq:cutsine}, where $\wradial=2\pi\times40.6$~Hz was held constant and all other parameters were varied. The fitted value of the reflection duration was $\Tref=0.9\pm0.1$~ms. This value is consistent with the expected duration given by $\sigma/v$, where $\sigma\approx6.6~\mu$m is the width of the wave packet and $v=3\hbar\kr/(2m)\approx7.5~$mms$^{-1}$ is the velocity of the wave packet just before reflection. The reflection time $\tref$ also agrees well with the expected value of 3.0~ms \cite{Sherson2012}.

\begin{figure}[]
	\includegraphics[width=8.6cm]{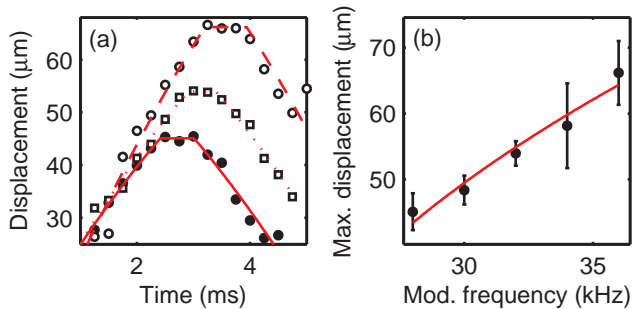}
	\caption{(Color online) Wave packet motion for different modulation frequencies. (a) Upper wave packet displacement as a function of hold time for $\numod$=28~kHz (filled circles, solid line), 32~kHz (open squares, dotted line) and 36~kHz (open circles, dashed line); the data for each frequency has been fitted with Eqn.~\eqref{eq:cutsine}. (b) Maximum wave packet displacement as a function of modulation frequency; data and errors were obtained from fits as in (a). Solid  line: fit to data using Eqn.~\eqref{eq:xmax}.}
	\label{fig:disp_v_freq}
\end{figure}

To further explore these reflections at the band gap and to analyze the  energy at which a wave packet is generated in the quasi-continuum, we investigated the wave packet's motion for several values of the modulation frequency. Figure~\ref{fig:disp_v_freq}~(a) shows the displacement of the upper wave packet as function of hold time for $\numod=28$, 32 and 36~kHz. Consistent with Fig.~\ref{fig:spectrum}~(a), higher modulation frequencies lead to a larger displacement at which the wave packet is reflected at the band gap. However, the modulation frequency  is limited by the width of the second excited band of about 7~kHz, since the excitation is a two step process passing through this band.

Figure~\ref{fig:disp_v_freq}~(b) shows the maximum turning point  position of the lower wave packet as a function of modulation frequency. This position can  be determined from energy conservation: Disregarding the perturbation of the harmonic potential due to the optical lattice, one can equate the modulation energy $2h\numod$ with the potential energy of the harmonic trap offset by the energy difference between the lowest and third excited bands, \mbox{$2h\numod = h\nu_{03} + \frac{1}{2}m\wradial^2 y^2$}. Solving for the position, we obtain
\begin{equation}
	y = \left[\frac{4h}{m \wradial^2}(\numod - \nu_{03}/2)\right]^\frac{1}{2}.
	\label{eq:xmax}
\end{equation}
This function was fitted to the data in Fig.~\ref{fig:disp_v_freq}~(b), yielding values of $\nu_{03}= 42.6\pm0.6$~kHz and 40.2$\pm0.8$~kHz for the upper and lower wave packet respectively. This is consistent with the value 41.1~kHz predicted from the band model; we attribute the asymmetry between upper and lower wave packets to anharmonicity in the combined potential, which is addressed in section~\ref{subsec:control}.

We do not observe effects due to collisions on the center of mass motion or the fitted widths of the wave packets as they move through the stationary condensate. This confirms that the density during a wave packet's traversal of the main cloud is sufficiently low that collisions do not play a significant role.

This detailed analysis of wave packet motion shows that the dynamical evolution of the system after modulation is well understood. The  improved data quality with respect to prior work~\cite{Sherson2012}, allows the observation of qualitatively new features such as the delay in the reflection region.

\section{De-excitation of wave packets}\label{sec:deexcitation}

\subsection{Introduction}\label{subsec:deexc-eff}

The wave packets represent a resource to produce states localized in a given lattice site in the lower bands of the combined trap: Following excitation to the quasi-continuum, the atoms can be de-excited using a second modulation pulse of the optical lattice. This was first demonstrated in Ref.~\cite{Sherson2012}. Here, we significantly extend this work showing that the frequency, amplitude, and pulse duration of the de-excitation can be tailored to adjust the number of atoms transferred to desired lattice sites. This allows us to populate several sites in the optical lattice, providing, for example, a path towards loading a quantum register. The method bears strong resemblance to the well-known pump-probe technique~\cite{Garraway1995}, which allows for the investigation of excited state dynamics and for the preparation of quantum states.

The starting point for these experiments is a wave packet moving in the quasi-continuum (see section~\ref{sec:exp-wp}) obtained by modulating the lattice at a frequency of 30~kHz and an amplitude of $\epsilon=0.15$ for a duration of 500~$\mu$s. By using a second modulation pulse, indicated as process (2) in Fig.~\ref{fig:spectrum}, this wave packet can be de-excited into lower states. Thus, the probe pulse  serves a dual purpose: It can be used to interrogate the evolution in the excited state, and it allows for the preparation of specific target states. 

\begin{figure}[tbp] \includegraphics[width=8.6cm]{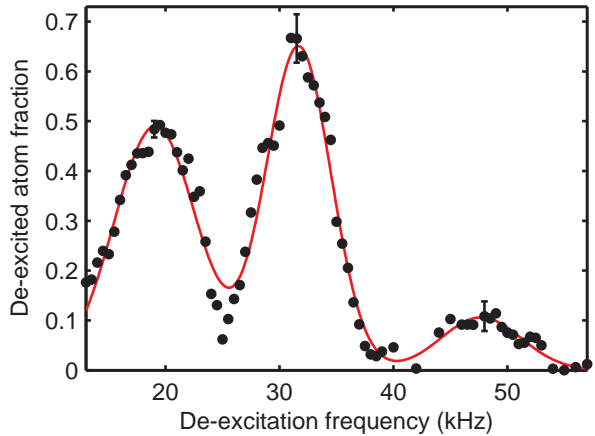}
	\caption{(Color online)	Fraction of the initial wave packet de-excited to localized states as a function of de-excitation frequency for a pulse amplitude of $\epsilon=0.15$ and a duration of 500~$\mu$s. The data shows the mean fraction in the upper localized state in a 2~kHz bin. 
	 The data has  been fitted with a triple Gaussian function, yielding resonances at 19.1, 31.7 and 47.7~kHz; these correspond to de-excitation into the second excited, first excited and lowest band respectively. The errorbars show the standard error of the mean for a given bin.}
	\label{fig:deexcitespectrum}
\end{figure}
We begin by investigating the modulation frequency dependence of the de-excitation process. The wave packet was allowed to evolve for 2~ms before applying a second modulation pulse to de-excite the atoms. Figure~\ref{fig:deexcitespectrum} shows the  fraction of atoms in the wave packet that were de-excited  as a function of modulation frequency. It shows three  resonances at $19.1\pm 0.1$, $31.7\pm0.07$, and $47.7\pm 0.7$~kHz, which correspond to transfer to the second excited, first excited and lowest bands respectively. This is consistent with the theoretically expected spacing shown in Fig.~\ref{fig:spectrum} of  6.3, 11.8 and 18.1 ~$\Erec$ corresponding to  17.3, 32.4 and 49.7~kHz respectively. 
To obtain the data shown in Fig.~\ref{fig:deexcitespectrum}, we modified the detection procedure to ensure that the de-excited atoms remained trapped in the lattice, while the remainder of the wave packets was removed~\footnote{The lattice depth was increased to $22.5~\Erec$ over 1~ms immediately following the de-excitation pulse. Subsequently, the magnetic trap was ramped down over 5~ms, and the atoms were held in the lattice for another 12~ms before taking an in-trap absorption image.   We have verified that a spurious two-step de-excitation to the first excited band while trying to de-excite to the second excited band is negligible for the data in Fig.~\ref{fig:deexcitespectrum}. This was achieved by reducing the lattice depth after de-excitation such that the second excited band was untrapped.}. This is most relevant for de-excitation into the second excited band, since part of this band is not trapped at a lattice depth of $16~\Erec$.

The de-excitation spectrum contains transitions that change the band index by $\Delta n = 1,2,3$. This does not violate the selection rule $\Delta n = 2$ introduced in section~\ref{sec:exp-wp}, since it only strictly applies  when the states involved are eigenstates of the parity operator. Nonetheless, the selection rule indicates that the transition rate for odd-numbered transitions will be reduced with respect to even-numbered ones. This is indeed the case in Fig.~\ref{fig:deexcitespectrum}, where the de-excitation fraction is reduced for the transition to the lowest band and the second excited band.

\begin{figure}[t] \includegraphics[width=8.6cm]{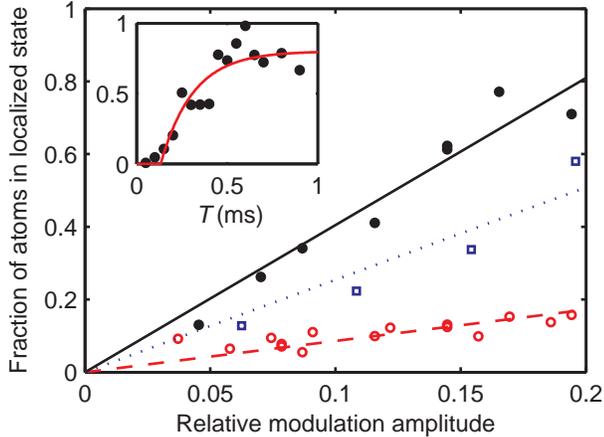}
	\caption{(Color online) Fraction of  atoms de-excited from the lower wave packet as a function of pulse amplitude. The experiment was performed for three de-excitation frequencies: 17~kHz (blue, open squares, dotted line), 34~kHz (black, filled circles, solid line), 49~kHz (red, open circles, dashed line). The data was corrected for off-resonant transfer to lower bands (see text). Linear fits were performed to guide the eye. Inset: Fraction of atoms in a localized state as a function of de-excitation pulse duration at a frequency of 32~kHz and $\epsilon= 0.17$. The data was fitted with an exponential saturation function with time constant $\tau=132~\mu$s. 
}
	\label{fig:deexcitevAmplitude}
\end{figure}

The fraction of de-excited atoms can also be controlled by the amplitude of the de-excitation pulse. We have investigated this dependence at modulation frequencies of 17, 34, and 49~kHz corresponding approximately to the resonances in the de-excitation spectrum shown above. Figure~\ref{fig:deexcitevAmplitude} shows the fraction of de-excited atoms as a function of the modulation amplitude. The maximal modulation amplitude in these experiments was limited to $\epsilon=0.2$, where the de-excitation begins to saturate. Above this value, re-excitation of atoms in localized states occurs, leading to non-linear behavior.
In the case of large amplitude transfer to the second excited band, we subtracted the signal of spurious atoms transferred to the first excited band. Figure~\ref{fig:deexcitevAmplitude} shows that the first excited band can be populated with high efficiency, i.e., a large fraction of the atoms in the initial wave packet can be de-excited to a localized state. In contrast, transfer into the lowest and the second excited band is limited, even at high modulation amplitudes. In general,  the modulation amplitude allows us to split the wave packet into a moving part and a number of localized parts in the bound  bands of the lattice.

Finally, the duration of the de-excitation modulation pulse can be used to tailor the de-excitation. The inset to Fig.~\ref{fig:deexcitevAmplitude} shows the number of de-excited atoms as a function of modulation pulse duration at a de-excitation frequency of 32~kHz. The de-excitation efficiency increases steadily until reaching saturation around $\SI{500}{\us}$. We obtain a time constant of $\tau=132~\mu$s from a fit to this data, which agrees well with the time constant obtained for the excitation process (see inset to Fig.~\ref{fig:pump-vs-freq}). This result validates the use of modulation pulses with a duration of $\SI{500}{\us}$ in both cases.

By tailoring both the amplitude and the pulse duration we can reach efficiencies approaching unity for the de-excitation to the first excited band, whereas transfer to the lowest and the second excited band is limited to 15\% and 50\%\ respectively. When coupling into the first excited band, the method thus allows for splitting the wave packet in an arbitrary ratio. Hence the wave packet can indeed serve as a resource to produce atoms localized in a given lattice site, providing, for example, a path towards loading an atomic quantum register. 

 In particular, the use of multiple de-excitation frequencies enables the controlled population of several, spatially separated, localized states, limited only by the spatial selectivity of the de-excitation pulse. Using this method, we have realized the splitting of a  propagating wave packet into two distinct localized states~\footnote{See Fig.~5 (d) in \cite{Sherson2012}; for this experiment, the initial wave packet was prepared using a modulation pulse at 33~kHz, and the de-excitation pulses were applied after 1.5~ms and 2~ms at frequencies of 26~kHz and 36~kHz respectively.}, thereby  demonstrating the ability to control the position and population of the localized wave packets.

\subsection{Spectroscopy of trap anharmonicity}\label{subsec:control}

In a first application of the technique, we show that the frequency dependence of the de-excitation allows for a  spectroscopic measurement of the band structure in the combined trap. This is in close analogy with the ability of other pump-probe schemes to characterize the final state after probing the system, e.g. a molecular ground state. Here, we measure the the spatial dependence of the resonant de-excitation frequency to characterize the potential with sufficient precision to detect distortions of the harmonic confinement.

To initiate the experiment, atoms were excited into the quasi-continuum by a modulation pulse of the optical lattice. In this case, the modulation frequency was swept from 27.5 to 28.5~kHz within 1.5~ms. This generated a temporally broad wave packet moving in the quasi-continuum. After a variable evolution time of the wave packet, a de-excitation pulse at a chosen modulation frequency was applied, and the magnetic trap was turned off. The de-excited atoms were then held in the lattice for a further 10~ms to allow atoms in the quasi-continuum to leave the detection region, and finally an absorption image was taken.
\begin{figure}[tbp] 
	\includegraphics[width=8.6cm]{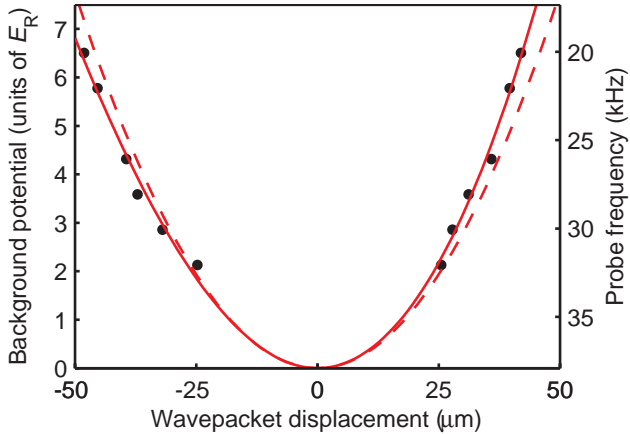}
	\caption{(Color online) Resonant de-excitation frequency as a function of position. The extracted background trapping potential (see text) is shown on the other vertical axis. The data was fitted with quadratic (dashed line) and cubic (solid line) functions to show the anharmonicity of the potential. The fitted offset ($c$) was subtracted from the data points and the fits.}
	\label{fig:anharmonicity}
\end{figure}
For each evolution time between 0.1 and 2~ms the de-excitation frequency was varied to obtain the resonant de-excitation frequency to the first excited band. The broadness of the initial wave packet was advantageous in this case because it reduced the number of necessary measurements.

Figure~\ref{fig:anharmonicity} shows the resonant de-excitation frequency as a function of the position of the de-excited atoms. The position for each evolution time was obtained from Gaussian fits to the absorption images. The link between the de-excitation frequency and the harmonic trapping potential can be seen directly in Fig.~\ref{fig:spectrum}~(a). The energy balance in this case shows that the resonant de-excitation frequency  is given by $\nudeex=2\numod-(E_\mathrm{bg}+E_{01})/h$.
where $E_{01}$ is the energy difference between the lowest and first excited bands, and $E_\mathrm{bg}$ corresponds to the background potential dominated by the magnetic trap. Since the first excited band is nearly flat, $E_{01}$ is approximately constant. 
The background potential  $E_\mathrm{bg}$ is therefore proportional to the de-excitation frequency. 

To illustrate our ability to resolve the anharmonicity of the trap, the data was fitted with a harmonic trapping potential $E_\mathrm{bg}=b' y^2+c$, and a harmonic potential plus a cubic term \mbox{$E_\mathrm{bg}=ay^3+by^2+c$,} where the offset $c$ was obtained from the cubic fit. For the quadratic fit, we obtain a $\chi^2$ value that is $9.3$ times larger than the value for the cubic fit, showing that the potential is indeed anharmonic. 

This anharmonicity is caused by a combination of factors including gravity, the inherent anharmonicity of the magnetic trap, and the imperfect alignment of the magnetic trap with the optical lattice.  Since the measured cubic component ($a$) is approximately $9.1$ times larger than the one expected under ideal conditions (i.e., perfect alignment), we attribute the anharmonicity shown in Fig.~\ref{fig:anharmonicity} primarily to the misalignment of the magnetic trap with the optical lattice. This observed anharmonicity can explain the asymmetry in the fitted value for $\nu_{03}$ in section~\ref{sec:wp-motion}.


\section{Multiply excited wave packets}\label{sec:up-excitation}

The ability to excite wave packets further into the quasi-continuum adds to the rich dynamical properties of this system. In  particular it allows for the creation of high momentum wave packets with potential applications in atom optics and atom interferometry. Such an excitation can be achieved by an appropriately chosen second modulation pulse of the lattice, which does not de-excite the atoms from a wave packet, but propels them further into the quasi-continuum. This further excitation is indicated as process (3) in Fig.~\ref{fig:spectrum}~(a). 

Since the wave packet moves in a quasi-continuum of states in an approximately harmonic trap, the excitation frequency spectrum is primarily dictated by a harmonic dispersion relation. Nonetheless, it remains useful to depict the process in a band structure picture, since it allows for an easy identification of the resonance energies. Figure~\ref{fig:spectrum}~(b) shows that the resonant excitation frequency depends on the momentum $\hbar q(t)$ of the primary wave packet. The interaction of atoms in the quasi-continuum states with the oscillating potential of the lattice can be visualized as absorption and emission of phonons carrying energy $\hbar\omega_{mod}$ and momentum $2\hbar \kr$. Consequently, absorption of such a phonon corresponds to a `vertical' transition into a higher band with band index $n_f = n_i + 2$. 

In our case, the primary wave packet is created close to the crossing between the third and fourth excited bands, and during its evolution it propagates in the third excited band. A further excitation at small initial $\hbar q$ hence transfers the wave packet to the crossing between the fifth and sixth excited bands, whereas later excitation at larger $\hbar q$ transfers the wave packet to the fifth excited band only. 

\begin{figure}[tbp]
	\includegraphics[width=8.6cm]{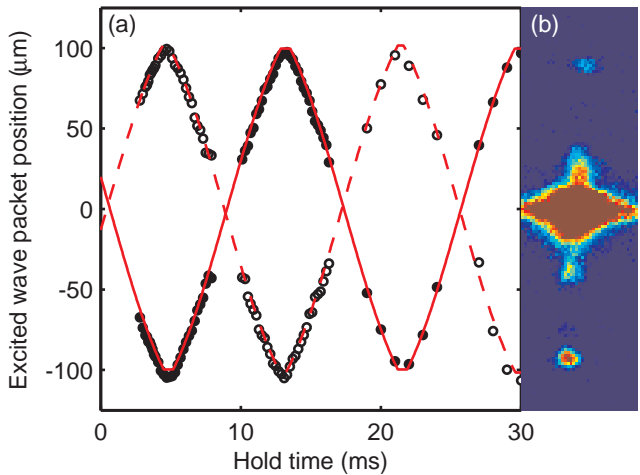}
	\caption{(Color online) Motion of multiply excited wave packets. (a) Lower wave packet data and fit: filled circles, solid line;  upper wave packet data and fit: open circles, dashed line. (b) Absorption image showing the multiply excited and the initial wave packet at a hold time of 4.5 ms. 
	}
	\label{fig:h}
\end{figure}

Figure~\ref{fig:h} shows the motion of such multiply excited wave packets. In this realization, the primary wave packets were created by modulating the lattice at 31~kHz, and after a delay of 2~ms a further excitation pulse at 52~kHz was applied. This results in a second transfer of  $2\hbar \kr$, yielding a doubly excited wave packet and hence a significantly larger maximal displacement during the following oscillation. The displacement of the wave packets was fitted with Eqn.~\eqref{eq:cutsine}. Due to the high momentum, the reflection duration $\Tref$ becomes small compared to the period of the oscillation, leading to a near sawtooth appearance. Figure~\ref{fig:h}~(b) shows an absorption image taken after a 4.5~ms hold time close to the wave packet's turning point at a displacement of $\approx 100~\mu$m. Both the primary wave packets and the multiply excited wave packets are visible.  

The observed displacement is consistent with the energy spectrum in Fig.~\ref{fig:spectrum}~(a), since the band gap is reached at larger displacements for multiply excited wave packets. Similarly, in momentum space (Fig.~\ref{fig:spectrum}~(b)), it takes a longer time for the wave packet to `roll down' multiple momentum branches under the influence of gravity before being reflected at the gap between the second and third excited bands. 
\begin{figure}
	 \includegraphics[width=8.6cm]{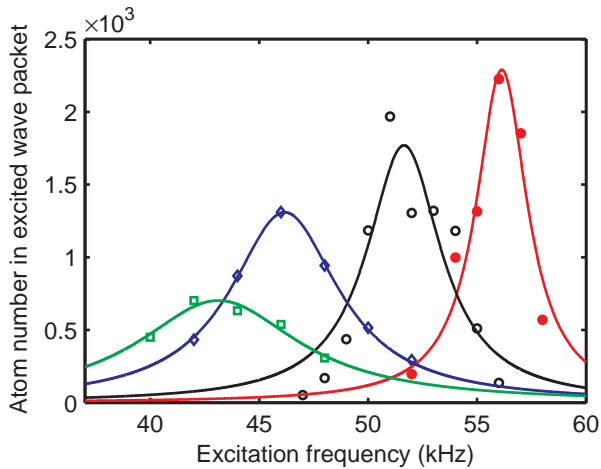}
	\caption{(Color online) Atom number in the highly excited wave packet as a function of the excitation frequency for four values of the delay between the modulation pulses: 0.6 ms (red filled circles), 2.0 ms (black circles), 2.9 ms (blue diamonds) and 3.3 ms (green squares). Solid lines indicate Lorentzian fits to each data set. The shift of the resonance frequency and the broadening of the excitation profile can be understood by considering the movement of the wave packet within the band structure.}
	\label{fig:mutiexcite}
\end{figure}

The resonant frequency for the second excitation depends on the momentum $\hbar q(t)$ of the primary wave packet, as outlined above. To map out this dependence, the frequency of the excitation pulse was varied for several values of the delay between the two modulation pulses. Figure~\ref{fig:mutiexcite} shows the atom number in the excited wave packet as function of excitation frequency. Each of the data sets was fitted with a Lorentzian to extract information about the resonance. The center frequency of the excitation profile  shifts down as the delay is increased, and at the same time the width of the resonance broadens.

\begin{figure}
	 \includegraphics[width=8.6cm]{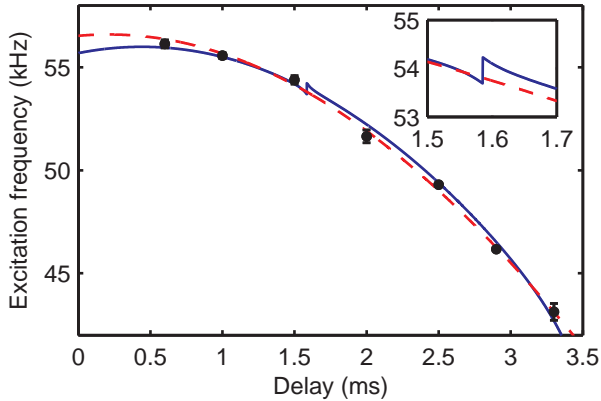}
	\caption{(Color online) Resonant transition frequency for the second excitation as a function of the delay between  modulation  pulses. The red dashed line is a fit to the data using Eqn.~\eqref{eq:i1}. The blue solid line is a  fit to a model based on a numerical calculation  of the band structure for the combined trap. It shows a small kink at the gap between the third and fourth excited bands, which is not resolved in the data; a close-up of this region is shown in the inset. The error bars correspond to the uncertainty of the resonant frequency extracted from the Lorentzian fits in Fig.~\ref{fig:mutiexcite}.}
	\label{fig:highfrequencyshift}
\end{figure}

 The shift of the resonance can be analyzed analytically for a free particle in a harmonic potential and numerically for a single particle in a band structure. The momentum of an atom in a given band is $q\hbar\kr$, and through the excitation it gains an additional $2\hbar\kr$ in momentum. For a free particle, the energy gained by an atom is thus given by
\begin{eqnarray}
 \frac{E^\prime - E}{\Erec}&=&\left( q + 2 \right)^2 - q^2, \nonumber \\
 &=& 4 \left[ 1 + \qmax \cos\left(\wradial(t-t_0) \right) \right],
  \label{eq:i1}
\end{eqnarray}
where we have replaced the momentum $q$ by that of a particle oscillating in a harmonic trap. Figure~\ref{fig:highfrequencyshift} shows the resonant frequencies as a function of the delay time. By fitting Eqn.~\eqref{eq:i1} to this data, the values of $\qmax=4.16\pm0.01~\kr$ and the time offset $t_0=0.21\pm0.02$~ms were obtained. Alternatively, the data was fitted by using a single particle band structure calculation. In that case the energy was calculated numerically for each value of the momentum $\qmax \cos\left(\wradial(t-t_0) \right)$, again using $\qmax$ and $t_0$ as free parameters. Thus values of $\qmax=4.18\pm0.02~\kr$ and the time offset $t_0=0.44\pm0.04~$~ms were obtained. The resulting resonant frequency has a small kink at $1.5$~ms which originates from the small gap between the third and fourth excited bands. This confirms that a transition from the fourth to the sixth excited band is driven for $t<1.5$~ms and from the third to the fifth excited band for $t>1.5$~ms. The obtained values of $\qmax$ agree well with the expected value of $4.22~\kr$~\footnote{The theoretical start momentum was obtained by finding the $q$ value where twice the excitation frequency matches the gap between the zeroth and fourth excited bands. Since the transition occurs between the first and fifth Brillouin zones, 4$\kr$ should be added to this value.}. The time offset reflects the time it takes to further excite atoms from the primary wave packet; the offset differs for the two methods, due to the influence of the lattice on the dispersion relation. Nonetheless, the reasonable agreement of both methods confirms the small influence of the lattice on the motion of the multiply excited wave packets.
 
These models also explain the broadening of the resonance. Since the primary wave packet  shows very little spreading~\cite{Sherson2012} during its evolution, its spatial width corresponds to a fixed temporal width with respect to the time delay between the excitation pulses. Due to the slope of the transition frequency in Fig.~\ref{fig:highfrequencyshift}, this corresponds to a narrow frequency distribution at short delays and a wide distribution at long delays. This trend is in agreement with the profiles shown in Fig.~\ref{fig:mutiexcite}.

The analysis above shows that an excellent understanding of the dynamics of multiply excited wave packets has been obtained. This method extends our ability to utilize wave packets in a number of ways. Since the second excitation occurs in the quasi-continuum, it is not sensitive to the particular lattice depth and can thus serve as a method to detect the momentum of the primary wave packet even when it is overlapped with the main cloud. In combination with the de-excitation technique, it allows for a controlled transport of atoms into distant localized states, not reachable by the primary wave packets. 

Finally, this technique in principle enables creation of very fast wave packets using moderate modulation frequencies. This can be achieved by applying an excitation pulse at an evolution time when the wave packet traverses the fourth or third excited band. In position space, this corresponds to a jump up the slope of the potential. Wave packets can thus achieve substantial velocity, depending on how many phonons they have absorbed. These high momentum wave packets have potential applications in atom optics and atom interferometry~\cite{Denschlag2002,PhysRevLett.100.180405,Clade2009,Park2012}.

\section{Conclusion and outlook}

In this paper, we have presented a detailed analysis of the creation, evolution and manipulation of wave packets in the combined potential of an optical lattice and a  magnetic trap. Our analysis of the production of wave packets shows that the lattice modulation technique and the two step excitation process are well understood, and wave packets can be reliably created as a resource for further experiments. By monitoring the wave packet motion, a detailed understanding of the dynamical evolution of the system has also been obtained. 

We  demonstrated the controlled de-excitation of atoms from the wave packets to the available bands of the combined trap. This confirms that lattice modulation can act as a beam splitter for matter wave packets with the ability to selectively address localized states in the available lower bands. Since these states lie well outside the region of the initial BEC, the bands at that location can be populated with high purity. 
In close analogy to the well-known method of pump-probe spectroscopy, the resonant frequency of the de-excitation pulse was used to perform spectroscopy of the anharmonicity of the trap. 
In addition, we have shown that multiple localized states can be created from a single wave packet.

Finally, it was shown that lattice modulation can be used to excite matter wave packets to higher momenta. This technique enables the creation of very fast wave packets using moderate modulation frequencies. These high momentum wave packets have potential applications in atom optics and atom interferometry.

Within future work it will be of particular interest to study the coherence properties of each individual step and to which extent coherent control techniques~\cite{Branderhorst2008} can be utilized. This will open up the possibility of creating superpositions of qubit states in, e.g., the lowest and first excited bands by modulating at multiple frequencies simultaneously. The combination of quantum degenerate atoms and optical lattices has proved useful for storing information in the form of qubits~\cite{Weitenberg2011}. The methods proposed here may provide future means to solve the outstanding challenge of transporting atoms between lattice sites in a controlled fashion.

\section{Acknowledgments}

We thank the Danish National Research Foundation and the Lundbeck Foundation for support. J.F.S. acknowledges support from the Danish Council for Independent Research.

%

\end{document}